\documentclass{MYws-ijmpa}
\usepackage[super,compress]{cite}
\usepackage{graphicx}
\usepackage{url,hyperref}
\usepackage{bbm}
\usepackage{xcolor}
\usepackage{booktabs} 
\usepackage{tabularx} 
\usepackage{longtable} 
\usepackage{array} 
\newcolumntype{L}{>{$}l<{$}} 

\newcommand{\V}[1]{\ensuremath{V_{#1}^{}}} 
\usepackage{suffix}
\WithSuffix\newcommand\V*[1]{\ensuremath{V_{#1}^*}}
\newcommand{\hc}{\mathrm{h.c.}}
\newcommand{\GeV} {\,\text{GeV}}
\newcommand{\TeV} {\,\text{TeV}}

\newcommand{\CHqone}{C^{(1)}_{Hq}}
\newcommand{\CHqonebracket}{\left[\CHqone\right]}
\newcommand{\CHqthree}{C^{(3)}_{Hq}}
\newcommand{\CHqthreebracket}{\left[\CHqthree\right]}
\newcommand{\CHud}{C_{Hud}}
\newcommand{\CHudbracket}{\left[\CHud\right]}

\newcommand{\QuantumNumbers}[3]{(\mathbf{#1}, \mathbf{#2}, #3)}

\def\PD      {\ensuremath{\mathrm{D}}\xspace}        
\def\Dbar    {\kern 0.2em\overline{\kern -0.2em \PD}{}\xspace}

\def\PK      {\ensuremath{K}\xspace} 
\def\Kbar  {\kern 0.2em\overline{\kern -0.2em \PK}{}\xspace}

\def\PB      {\ensuremath{B}\xspace}         
\def\Bbar    {\ensuremath{\kern 0.18em\overline{\kern -0.18em \PB}{}}\xspace}

\newcommand{\ie}{{\em i.e.}}

\begin{document}
\markboth{Teppei Kitahara}{Theoretical point of view on CAA}

%
\catchline{}{}{}{}{}
%

\title{Theoretical point of view on Cabibbo angle anomaly} 

\author{Teppei Kitahara}

\address{Department of Physics, Graduate School of Science,
Chiba University, Chiba 263-8522, Japan\\ 
Kobayashi-Maskawa Institute for the Origin of Particles and the Universe,\\
Nagoya University, Nagoya 464-8602, Japan\\
CAS Key Laboratory of Theoretical Physics, Institute of Theoretical Physics,\\
Chinese Academy of Sciences, Beijing 100190, China\\
kitahara@chiba-u.jp}

\maketitle


\begin{abstract}
We present the current situation of the determinations of the first-row CKM components and show the Cabibbo angle anomaly corresponding to a deficit in the first-row CKM unitarity condition at the $3\sigma$ level. 
In this contribution, we show two new physics interpretations: heavy vector-like quark models and a MeV scale sterile neutrino models. 
The super tau-charm facility will directly probe the other CKM unitarity conditions related to $V_{cd}$.

\end{abstract}

\ccode{Report number: CHIBA-EP-265}


\section{Introduction}

\renewcommand{\thefootnote}{\fnsymbol{footnote}}
\footnotetext[0]{Presented at the 2024 International Workshop on Future Tau Charm Facilities (FTCF2024), Hefei, China,
14-18 January, 2024. The original talk title is ``Anomalies in the charm-strange sector.''}

Precision experiments of flavor physics are becoming increasingly important in indirectly discovering physics beyond the Standard Model (SM). 
In particular, there is the so-called Cabibbo angle anomaly (CAA)~\cite{Belfatto:2019swo,Grossman:2019bzp,Seng:2020wjq,Coutinho:2019aiy,Manzari:2020eum,Crivellin:2020ebi,Kirk:2020wdk,Crivellin:2020lzu,Capdevila:2020rrl,Belfatto:2021jhf,Crivellin:2021njn} with a significance of currently around the $3\sigma$ level~\cite{Bryman:2021teu,Cirigliano:2022yyo,Seng:2022ufm,Crivellin:2022rhw}.
The CAA consists of two tensions related to the determination of the Cabibbo angle: First, the different determinations of $|V_{us}|$ from $K_{\mu 2}$, $K_{\ell3}$, and $\tau$ decays disagree at the $3\sigma$ level. 
Second, using the average of these results in combination with $\beta$ decays, a deficit in first-row Cabibbo-Kobayashi-Maskawa (CKM) unitarity appears with significance at the $3\sigma$ level.
Intriguingly, it is known that both discrepancies could be explained via a modified $W$ coupling to quarks. 

Importantly, due to $SU(2)_L$ invariance, such a modified $W$ coupling to quarks in general leads to modified $Z$-quark-quark couplings as well, that enter electroweak precision observables (EWPO), affect low-energy parity violation and can give contributions to the flavor-changing-neutral-current (FCNC) processes. Therefore, a global fit is required to consistently assess the agreement of a specific New Physics (NP) scenario with data. The necessity of such a combined analysis becomes even more obvious when considering a UV complete model that can generate modified $W$ couplings to quarks. 

We will also discuss a light new physics interpretation of the CAA by considering the sterile neutrino models.

\section{Current status of Cabibbo angle anomaly}

\begin{figure}[t]
\centerline{\includegraphics[width=7cm]{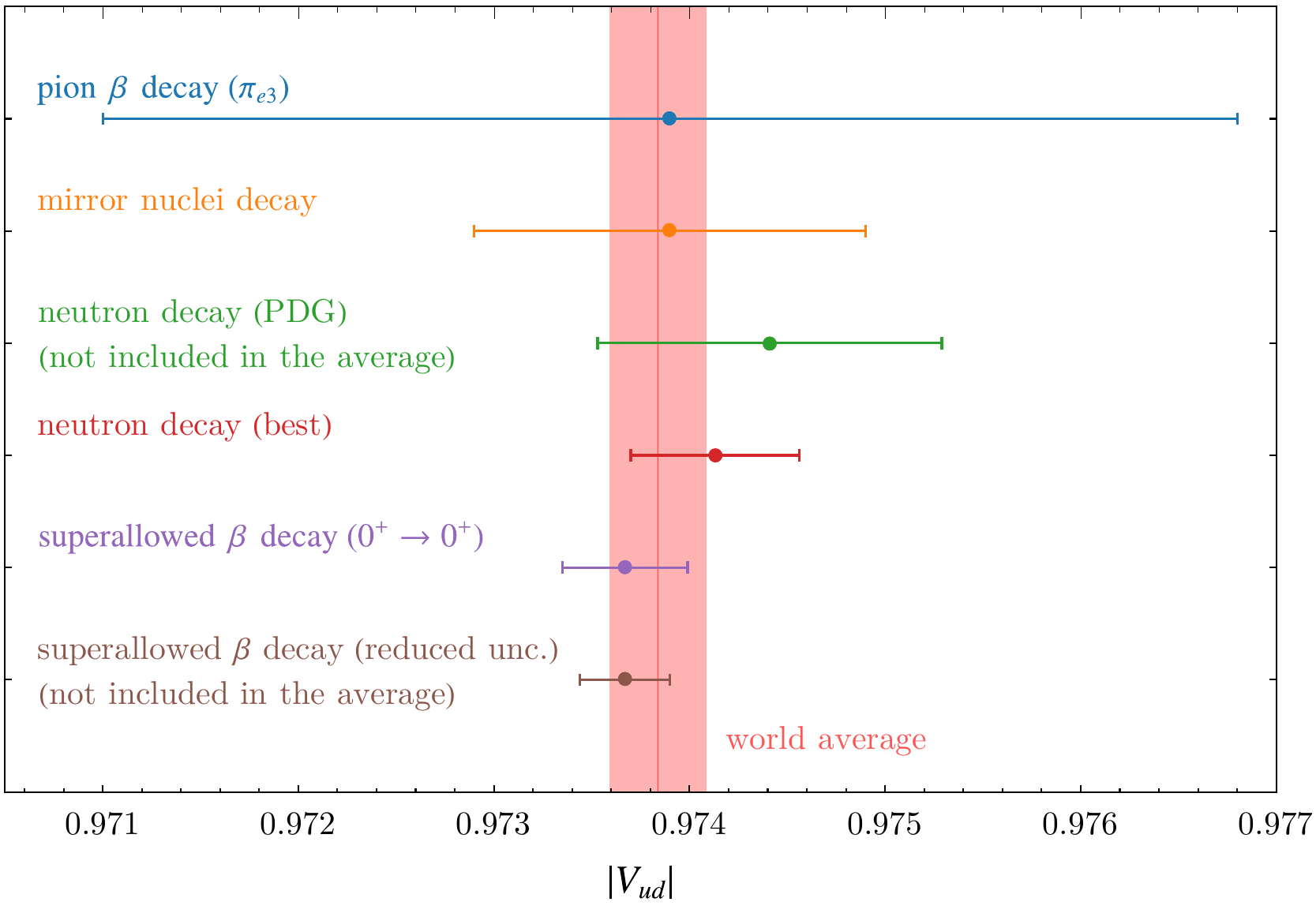}\quad
\includegraphics[width=7cm]{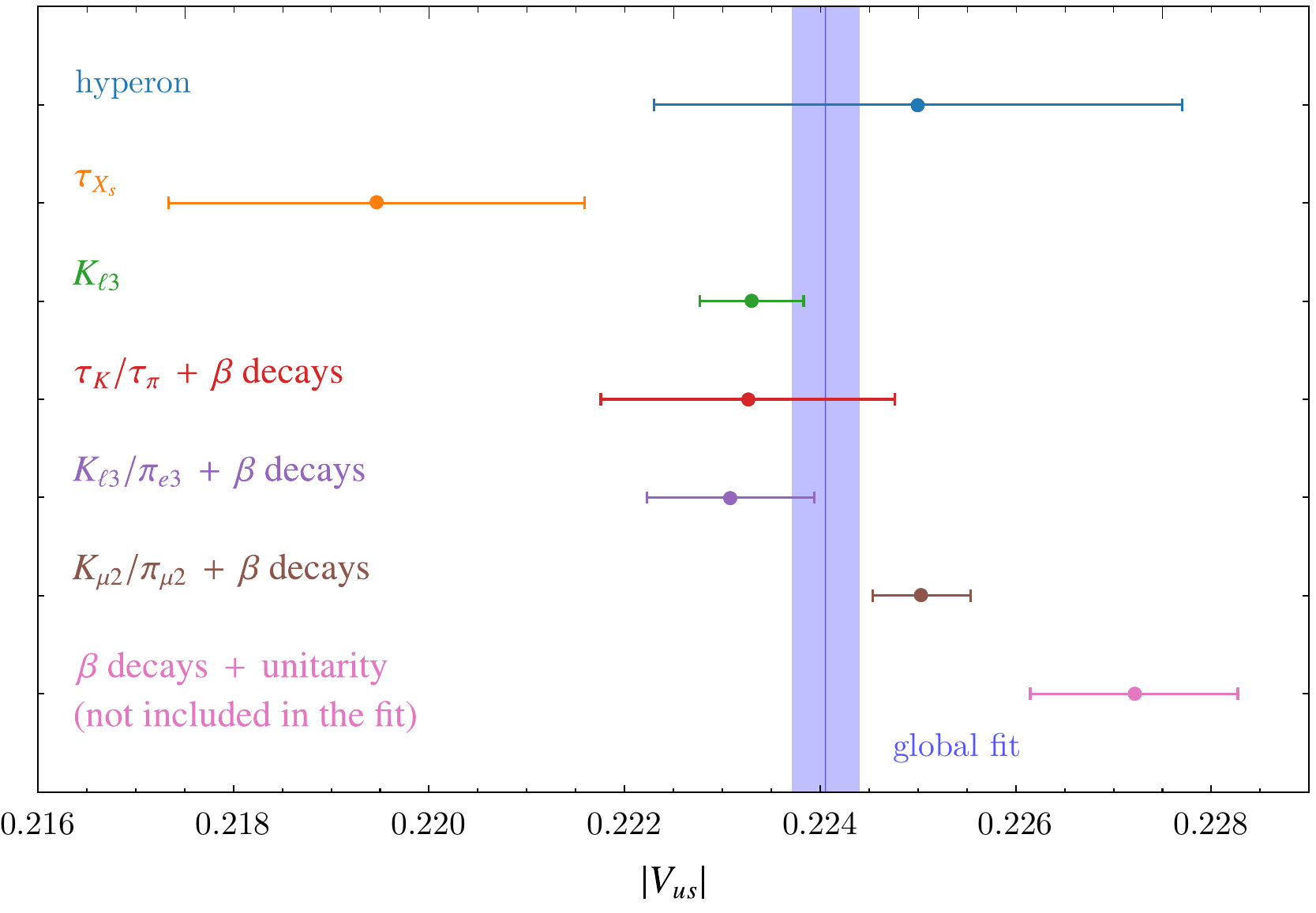}}
\caption{(Left) Summary of the determinations of $|V_{ud}|$ from the various types of $\beta$ decays \cite{Crivellin:2022rhw}. The red band represents the world average in Eq.~\eqref{eq:betaWA}.
The details of the extractions from
 neutron decay (best) and the super-allowed $\beta$ decays (reduced uncertainty) are given in the main text. Note that the pion $\beta$ decay, even though it is currently not competitive, is theoretically clean and will be strikingly improved by the PIONEER experiment~\cite{PIONEER:2022alm}. (Right) Summary of the determinations of $|V_{us}|$ from various processes \cite{Crivellin:2022rhw}. The global fit value of $|V_{us}|$ is obtained in Eq.~\eqref{eq:Vusglobal}.}
 \label{fig:beta_status}
\end{figure}

Here, we briefly summarize the current situations of the $|V_{ud}|$ and $|V_{us}|$ determinations shown in Fig.~\ref{fig:beta_status}.

First, the CKM element $|V_{ud}|$ 
can be determined from various types of $\beta$ decays.
The latest determinations are $|\V{ud}|_{0^+\to 0^+}=0.97367(32)$ from the super-allowed $0^+ \to 0^+$ nuclear $\beta$ decay~\cite{Hardy:2020qwl,Cirigliano:2022yyo}, 
$|\V{ud}|_{n{\rm (PDG)}} = 0.97441(88)$ 
from the neutron decay~\cite{ParticleDataGroup:2022pth},
$|\V{ud}|_{\rm mirror} = 0.9739(10)$ 
from $\beta$ transitions of the mirror nuclei \cite{Hayen:2020cxh}, and 
$|\V{ud}|_{\pi_{e 3}} = 0.9739(29)$
from the pion $\beta$  decay ($\pi^+ \to \pi^0 e^+ \nu;~\pi_{e3}$)~\cite{Feng:2020zdc,Cirigliano:2022yyo}.
In these determinations, we use an estimation of Ref.~\cite{Cirigliano:2022yyo} for universal nuclear-independent radiative corrections from $\gamma W$-box diagrams $\Delta^V_R$ \cite{Marciano:2005ec}.\footnote{%
A new lattice calculation  derives $\Delta_R^V=0.02439(19)$ \cite{Ma:2023kfr}. 
If we adopt this result for the global fit of the super-allowed nuclear $\beta$ decays,
significance of the tension is reduced by about 0.5$\sigma$.} 
For the neutron decay,
it is known that the uncertainty of $|\V{ud}|_{n{\rm (PDG)}}$ is inflated by scale factors that come from inconsistencies in the data. By using the single most precise result for the neutron lifetime $\tau_n$ \cite{UCNt:2021pcg} and the nucleon isovector axial charge $g_A/g_V$ \cite{Markisch:2018ndu}, a better determination of  $|V_{ud}|_{n{\rm (best)}} = 0.97413(43)$ is possible~\cite{Cirigliano:2022yyo}. 
Combining $|V_{ud}|_{0^+\to0^+}$, $|V_{ud}|_{n{\rm (best)}} $, $|\V{ud}|_{\rm mirror}$, and $|\V{ud}|_{\pi_{e 3}}$, we obtain a weighted average of
\begin{equation}
|V_{ud}|_{\beta}= 0.973\,84(25)\,.
\label{eq:betaWA}
\end{equation}
Note that the $|V_{ud}|$ determination is predominated by super-allowed $\beta$ decays where the largest uncertainty 
comes from nuclear-structure (NS) dependent radiative corrections~\cite{Gorchtein:2018fxl}, encoded in $\delta_{{\rm NS},E}$ in Ref.~\cite{Hardy:2020qwl}. Unfortunately, precise estimations of $\delta_{{\rm NS},E}$ are difficult~\cite{Seng:2022cnq} but the current value is considered to be very conservative~\cite{Gorchtein:2018fxl}.
Recent progress for the NS-dependent radiative corrections has been made in Refs.~\cite{Cirigliano:2024msg,Cirigliano:2024rfk,Gennari:2024sbn}.

Next, the matrix element $|V_{us}|$ can be determined from semi-leptonic decays of kaons and hyperons and from inclusive hadronic $\tau$ decays. By comparing theoretical predictions with data of the semi-leptonic kaon decays $K_{S,L} \to \pi^- \ell^+ \,\nu$ and $K^+ \to \pi^0 \ell^+ \,\nu$ with $\ell = e, \mu$ (labeled $K_{\ell 3}$), one can obtain~\cite{Cirigliano:2022yyo}, 
$
    |V_{us}|_{K_{\ell 3}} = 0.223\, 30(53),
$
where the latest evaluations of the long-distance electromagnetic (EM)  correction \cite{Seng:2021boy,Seng:2021wcf,Seng:2021nar,Seng:2022wcw}, the strong isospin-breaking
correction \cite{Cirigliano:2022yyo},
and the recent $K_S$ data from the KLOE-2 collaboration 
\cite{KLOE-2:2019rev,KLOE-2:2022dot} are used.
Beyond kaons, one can also use the hyperon semi-leptonic decays, $(\Lambda \to p, \Sigma \to n, \Xi \to \Lambda, \Xi \to \Sigma ) \,\ell\, \overline\nu$, which however lead to a slightly different yet less precise value
$
|V_{us}|_{\rm hyperon} = 0.2250(27)
$ \cite{ParticleDataGroup:2022pth}.
Inclusive hadronic $\tau$ decays also provide an opportunity to extract the matrix element $|V_{us}|$  by separating the strange and non-strange hadronic states.
Two representative determinations are reported: $ |V_{us}|_{\rm HFLAV}  = 0.2184 (21)$~\cite{Gamiz:2006xx,HFLAV:2022pwe,Lusiani:2022} and $ |V_{us}|_{\rm OPE+lattice}  = 0.2212(23)$~\cite{Hudspith:2017vew,Maltman:2019xeh}.
The former is based on the conventional operator product expansion (OPE) using the vacuum saturation approximation~\cite{Pich:1999hc}, while the latter is based on improved OPE series by fitting the lattice result~\cite{RBC:2010qam}.
Although they almost agree, 
there is no common consensus on which value, $|\V{us}|_{\rm HFLAV}$ or $|V_{us}|_{\rm lattice}$, to use. Accordingly, we perform a weighted average of the two values
$
 |V_{us}|_{\tau_{X_s}} =  0.2195(21) .
$
By using these $|\V{us}|$ determinations, we obtain  
a weighted average of $|V_{us}|_{K_{\ell 3}}$, $|V_{us}|_{\rm hyperon}$, and $|V_{us}|_{\tau_{X_s}}$,
\begin{align}
 |V_{us}|_{K,\tau,\Lambda} = 0.223\,14(51)\,.
\end{align}

We summarize the determinations of $|V_{us}|$ from various observables in the right panel of Fig.~\ref{fig:beta_status}. 
There, the blue band represents the global fit of $|V_{us}|$ in Eq.~\eqref{eq:Vusglobal} in which the CKM unitarity condition is not included. It is shown that $|V_{us}|_{\tau_{X_s}}$ (orange bar) is a little bit smaller than the other determinations; $3.3\sigma,\,2.6\sigma,\,1.8\sigma$ discrepancies 
by comparing to $\beta$ decays with unitarity (magenta), 
$K_{\mu 2}/\pi_{\mu 2}$ with $\beta$ decays (brown), and $K_{\ell 3}$ (green), respectively.

Third, the ratio $|V_{us}/V_{ud}|$ can be extracted from the several ratios of leptonic decay rates of kaon, pion  and $\tau$ leptons. The leptonic kaon-decay rate over the pion one, $K_{\mu2}/\pi_{\mu2}=\Gamma(K^+ \to \mu^+\nu)/\Gamma(\pi^+ \to \mu^+  \nu)$  provides
$
    \left| V_{us}/V_{ud} \right|_{K_{\mu2}/\pi_{\mu2}} = 0.231\,08(51)
$ \cite{Moulson:CKM21},
where the latest evaluation of the long-distance EM and strong isospin-breaking corrections \cite{Giusti:2017dwk,DiCarlo:2019thl} is used.
Furthermore, the exclusive $\tau$-decay ratio $\Gamma(\tau \to K^- \nu)/\Gamma(\tau \to \pi^- \nu)$ (labelled by $\tau_K/\tau_\pi$) provides 
$
    \left|V_{us}/V_{ud}\right|_{\tau_K/\tau_\pi} = 0.2293(15) 
$ \cite{Lusiani:2022wcp}.
In addition, it is recently pointed out in Ref.~\cite{Czarnecki:2019iwz} that the semi-leptonic kaon-decay rate over the pion $\beta$ decay, 
$\Gamma(K \to \pi \ell \nu)/\Gamma (\pi^+ \to \pi^0 e^+ \nu)$ (labelled by $K_{\ell 3}/\pi_{e3}$), provides
$
  \left|V_{us}/V_{ud}\right|_{K_{\ell 3}/\pi_{e3}} = 0.229\,08(88) 
$ \cite{Seng:2021nar}.
Again, we obtain a weighted average of $|\V{us}/\V{ud}|_{K_{\mu2}/\pi_{\mu2}}$, 
$|\V{us}/\V{ud}|_{\tau_K/\tau_\pi}$ and $|\V{us}/\V{ud}|_{K_{\ell 3}/\pi_{e3}}$,
\begin{align}
    \left|\frac{V_{us}}{V_{ud}}\right|_{\rm ratios} = 0.230\,47(43)\,.
\end{align}

\begin{figure}[t]
\centerline{\includegraphics[width=7cm]{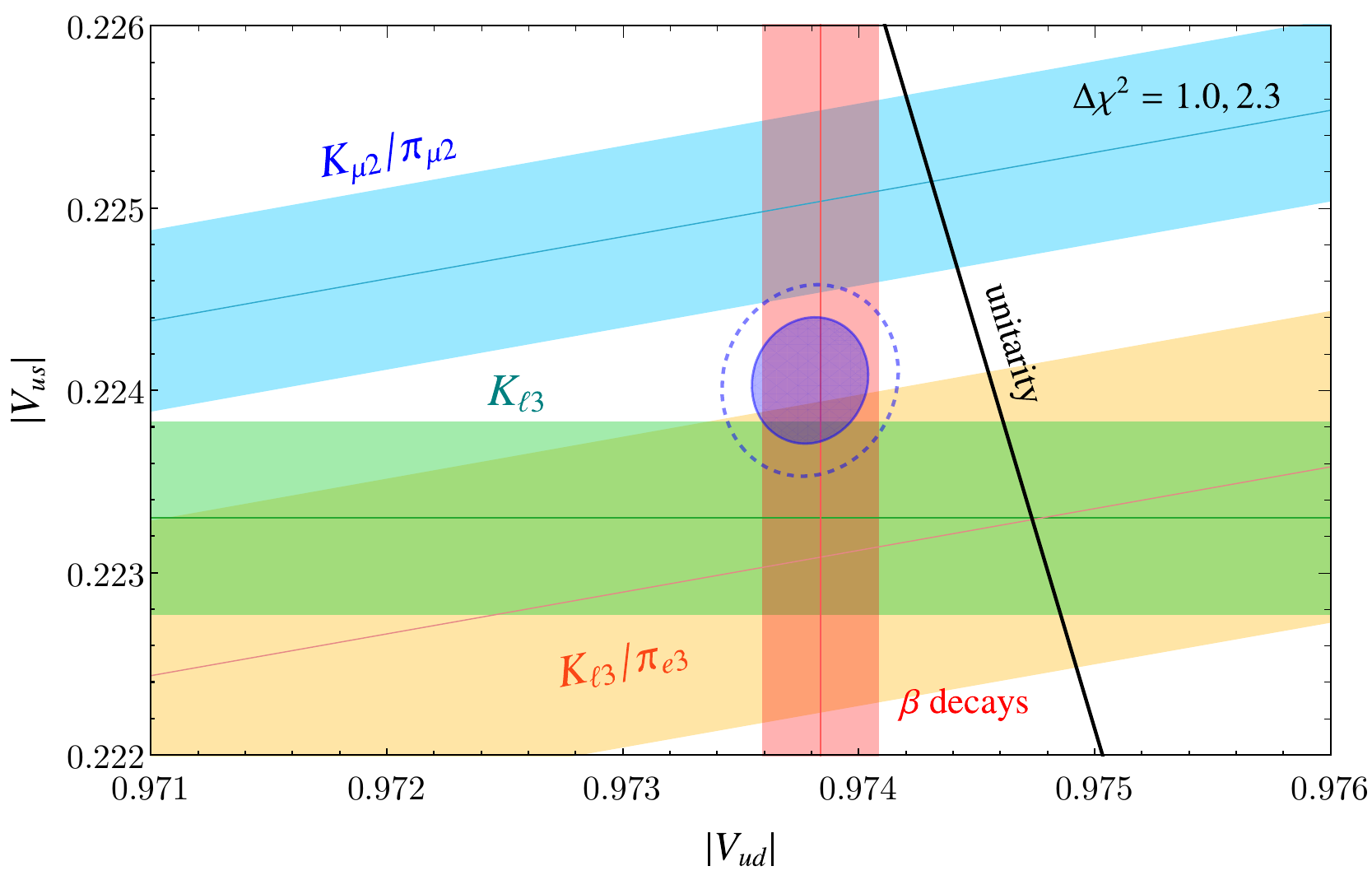}\quad
\includegraphics[width=7cm]{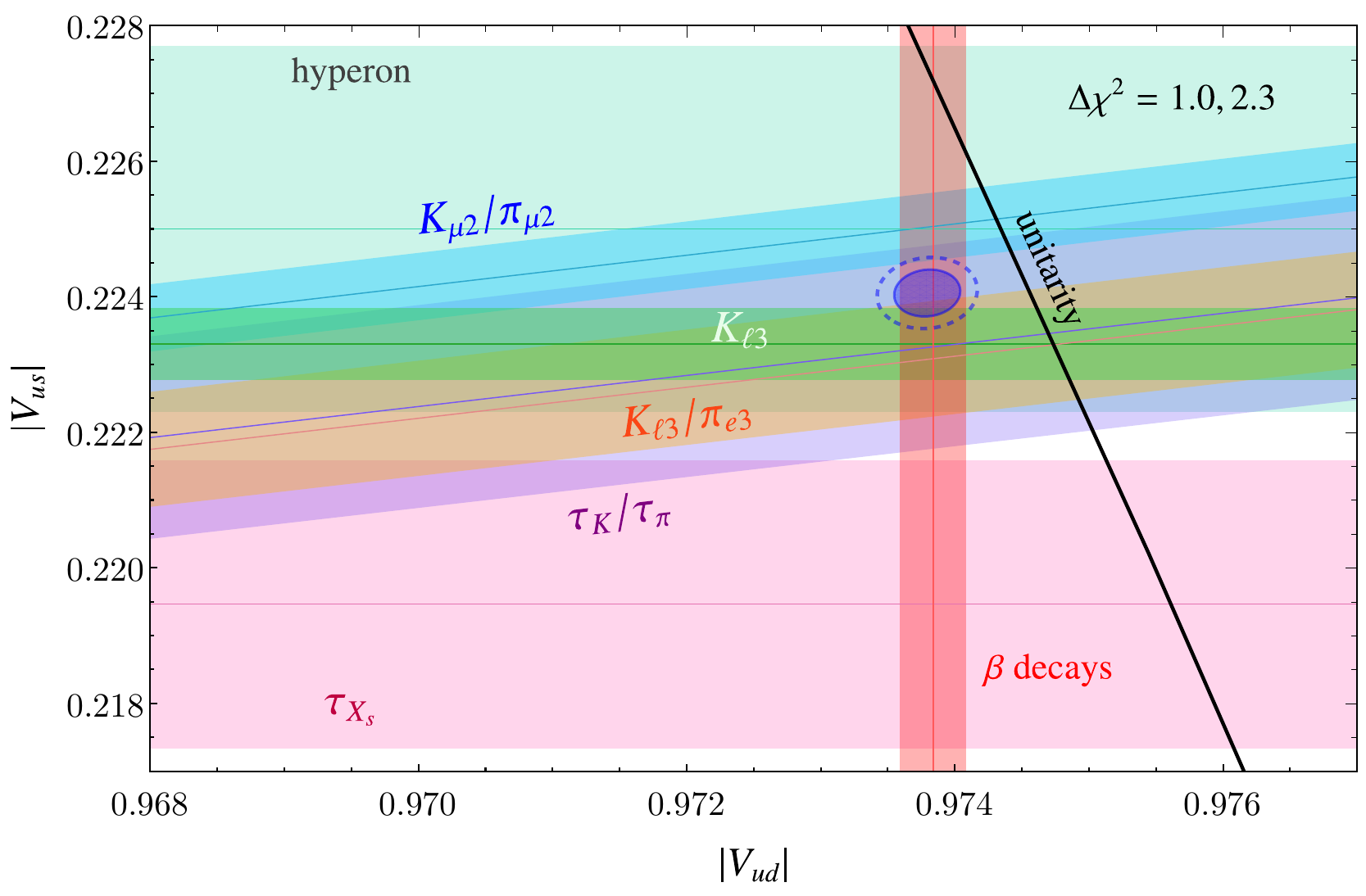}}
\caption{(Left) Global fit of all the available CKM determinations   
with $\Delta \chi^2 = 1$ (blue shaded) and $\Delta \chi^2 = 2.3$ (dashed circle) \cite{Crivellin:2022rhw}. Only the $1\sigma$ regions from $\beta$ decays, $K_{\ell 3}$, $K_{\mu 2}/\pi_{\mu 2}$ and $K_{\ell 3}/\pi_{e 3}$ observables are shown. The black line represents the unitarity condition. (Right) The $1\sigma$ regions of $|V_{us}|_{\rm hyperon} $, $ |V_{us}|_{\tau_{X_s}} $ and 
$|V_{us}/V_{ud}|_{\tau_K/\tau_\pi}$ observables are also shown \cite{Crivellin:2022rhw}.}
\label{fig:CAA_status}
\end{figure}

Finally, we perform a global analysis within the SM.
In Fig~\ref{fig:CAA_status}, 
the global fit result including $|V_{ud}|_{\beta}$, 
$ |V_{us}|_{K,\tau,\Lambda} $ and 
$|V_{us}/V_{ud}|_{\rm ratios} $ is shown by the blue circles.
In the left panel, 
only $\beta$ decays, $K_{\ell 3}$, $K_{\mu 2}/\pi_{\mu 2}$ and $K_{\ell 3}/\pi_{e 3}$ are displayed (but all data are included in the global fit), 
while the right panel shows all the data.
The black line stands for the unitarity condition: $|V_{ud}|^2+|V_{us}|^2 + |V_{ub}|^2=1$ with $|V_{ub}| \approx 0.00377$ \cite{Charles:2004jd}.
The blue shaded circle corresponds to $\Delta \chi^2 \leq 1$, while the dashed circle is $\Delta \chi^2 = 2.3$.
In the $\chi^2$ analysis, we included a correlation between $K_{\ell 3}$ and  $K_{\ell 3}/\pi_{e 3}$ because they share the same kaon data and common form factor $f_{+}^{K^0 \to \pi^-}(0)$. We set $100\%$ correlation for these common uncertainties.

Our global fit results are
\begin{align}
    |V_{ud}|_{\text{global}} & = 0.973\,79 (25)\,,
    \label{eq:Vudglobal}\\
    |V_{us}|_{\text{global}} & = 0.224\,05 (35)\,,
    \label{eq:Vusglobal}
\end{align}
and
\begin{align}
    \Delta_{\text{CKM}}^{\text{global}} & \equiv | V_{ud} |^2_{\text{global}} + | V_{us}|^2_{\text{global}} + |V_{ub}|^2 -1  = -0.001\,51(53)\,,
    \label{eq:CAAglobal}
\end{align}
with a $ |V_{ud}|_{\text{global}}$--$|V_{us}|_{\text{global}}$ correlation of $0.09$. This $\Delta_{\text{CKM}}^{\text{global}}$
 implies $- 2.8\sigma$ deviation from the unitarity condition of the CKM matrix.
 
One should note that the value of $ \Delta_{\text{CKM}}^{\text{global}}$ is also sensitive to the data of the neutron lifetime. The above result is based on the bottle ultracold neutron (UCN) lifetime data
$\tau_n^{\rm bottle} =877.75(36)$\,sec~\cite{UCNt:2021pcg}.
On the other hand, if instead the in-beam neutron lifetime data $\tau_n^{\rm beam}=887.7(2.2)$\,sec
  \cite{Yue:2013qrc} is used,
  $  \Delta_{\text{CKM}}^{\text{global}}$ is slightly changed  \cite{Kitahara:2023xab}
\begin{align}
    \Delta_{\text{CKM}}^{\text{global}} &   = -0.002\,34(62)\,,
    \label{eq:CAAglobalbeam}
\end{align}
which  
implies $- 3.8\sigma$ deviation from the unitarity condition.

Before closing this section, we give the status of the other unitarity tests:
\begin{align}
  \Delta_{\rm CKM}^{1^{\rm st}{\rm column}} &\equiv  |V_{ud}|^2 + |V_{cd}|^2 + |V_{td}|^2 - 1 = -0.0028(18)\,,\\
    \Delta_{\rm CKM}^{2^{\rm nd}{\rm row}} &\equiv  |V_{cd}|^2 + |V_{cs}|^2 + |V_{cb}|^2 - 1 = 0.0002^{+15}_{-13}\,,\\
      \Delta_{\rm CKM}^{1^{\rm st}\times 2^{\rm nd}{\rm row}} & \simeq  -  |V_{ud}| |V_{cd}| + |V_{us}||V_{cs}|   = 0.003(4)\,.
\end{align}
These quantities are still consistent with the CKM unitarity. 
The above uncertainties are dominated by $|\V{cd}|$, which is dominated by $D^+ \to \mu^+ \nu$ measurement~\cite{BESIII:2013iro}.
The super tau-charm facility (STCF) will provide $\mathcal{O}(0.1)\%$ statistical accuracy for the $D$-meson decays, which reduce the uncertainties of 
$|\V{cd}|$ and $|\V{cs}|$.
Therefore, it is expected that the STCF can probe these CKM unitarity tests.

\section{SMEFT fittings}

Here, we briefly summarize a situation of the Standard Model Effective Field Theory (SMEFT) interpretation of the CAA in the previous section.

We write the SMEFT Lagrangian as
\begin{align}
\mathcal{L}_\text{SMEFT} = \mathcal{L}_{\rm SM} + \sum_i C_i Q_i\,,
\end{align}
such that the SMEFT coefficients have dimensions of inverse mass squared. We use the Warsaw basis~\cite{Grzadkowski:2010es}, as well as the corresponding conventions, in which the operators generating modified gauge-boson couplings to quarks are given by
	\begin{align}
	\begin{aligned}
			Q_{H q}^{(1)ij} &= (H^{\dagger}i\overset{\leftrightarrow}{D_{\mu}}H)(\bar{q}_i\gamma^{\mu} P_L q_j)\,, &
			Q_{H q}^{(3)ij} &= (H^{\dagger}i\overset{\leftrightarrow}{D_{\mu}^I}H)(\bar{q}_i\tau^I\gamma^{\mu} P_L q_j)\,,\\
			Q_{H u}^{ij} &= (H^{\dagger}i\overset{\leftrightarrow}{D_{\mu}}H)(\bar{u}_i\gamma^{\mu}P_R u_j)\,, &
			Q_{H d}^{ij} &= (H^{\dagger}i\overset{\leftrightarrow}{D_{\mu}}H)(\bar{d}_i\gamma^{\mu}P_R d_j)\,,\\
			Q_{H ud}^{ij} &= i(\tilde{H}^{\dagger}D_{\mu}H)(\bar{u}_i\gamma^{\mu} P_R d_j)\,. & &
		\label{eq:HiggsqOperators}
		\end{aligned}
	\end{align}
We work in the down-basis such that CKM elements appear in transitions involving left-handed up-type quarks after the EW symmetry breaking. This means we write the left-handed quarks doublet as $q^T_i = \begin{pmatrix}(V^\dag u_{L})_{i} & d_{L, i} \end{pmatrix}$, where $V$ is the CKM matrix. With this conventions, the modified $W$ and $Z$ couplings are given by
\begin{equation}
\mathcal{L}_{W,Z} = 
\begin{aligned}[t]
&- \frac{g_2}{\sqrt{2}} W^+_\mu \, \bar{u}_i \gamma^\mu \left( \left[ V \cdot \left( \mathbbm{1} + v^2  \CHqthree \right) \right]_{ij} P_L + \frac{v^2}{2} \CHudbracket_{ij} P_R \right) d_j + \hc \\
&- \frac{g_2}{6 c_W} Z_\mu \, \bar{u}_i \gamma^\mu 
	\begin{aligned}[t]
	\Bigg( &\left[ (3-4s_W^2) \mathbbm{1} + 3 v^2 \, V \cdot \left\{ \CHqthree - \CHqone \right\} \cdot V^\dagger \right]_{ij} P_L \\
	- &\left[4s_W^2 \mathbbm{1} + 3 v^2 C_{Hu}\right]_{ij} P_R \Bigg) u_j
	\end{aligned} \\
&- \frac{g_2}{6 c_W} Z_\mu \, \bar{d}_i \gamma^\mu 
	\begin{aligned}[t]
 	\Bigg( &\left[ (2s_W^2-3) \mathbbm{1} + 3 v^2 \left\{ \CHqthree + \CHqone \right\} \right]_{ij} P_L \\
	+ &\left[ 2s_W^2 \mathbbm{1} + 3 v^2 C_{Hd}\right]_{ij} P_R \Bigg) d_j \,,
	\end{aligned}
\end{aligned}
\label{eq:WZCH}
\end{equation}
with $v \simeq {246}{\GeV}$.

\begin{table}
\tbl{Best fit points, $\Delta \chi^2$ and pulls w.r.t.~the SM hypothesis for the various EFT scenarios \cite{Crivellin:2022rhw}. The best fit points are in units of $10^{-3} v^{-2}$.\label{tab:eft_best_fits}}
{\begin{tabular}{@{}LLLL@{}}
\toprule
\text{EFT Scenario} & \text{Best fit point} & - \Delta \chi^2 & \text{Pull} \\
\midrule 
\CHqthreebracket_{11} & -0.50 & 3.3 & {1.8}{\sigma}
\\
\CHqthreebracket_{11} = \CHqthreebracket_{22} & -0.27 & 1.1 & {1.1}{\sigma}
\\
\CHqthreebracket_{11} = \CHqonebracket_{11} & -0.55 & 3.7 & {1.9}{\sigma}
\\
\CHudbracket_{11} & -1.0 & 3.1 & {1.8}{\sigma}
\\
\CHudbracket_{12} & -2.0 & 7.4 & {2.7}{\sigma}
\\
\left(\CHudbracket_{11}, \CHudbracket_{12}\right) & (-1.4, -2.1) & 13 & {3.2}{\sigma} 
\\
\left(\CHqthreebracket_{11}, \CHudbracket_{12}\right) & (-0.43, -2.0) & 11 & {2.8}{\sigma}
\\
\left(\CHqthreebracket_{11}, \CHudbracket_{11}, \CHudbracket_{12}\right) & (0.27, -1.9, -2.4) & 16 & {2.9}{\sigma}
\\
\left(\CHqthreebracket_{11}, \CHqthreebracket_{22}, \CHudbracket_{11}, \CHudbracket_{12}\right) & (0.59, 0.76, -2.6, -2.5) & 17 & {2.9}{\sigma}
\\
\left(\CHqthreebracket_{11}, \CHqonebracket_{11}, \CHudbracket_{11}, \CHudbracket_{12}\right) & (0.29, 0.11, -2.0, -2.4) & 13 & {2.6}{\sigma}
\\
\bottomrule
\end{tabular}
}
\end{table}

In Table~\ref{tab:eft_best_fits}, the best fits of the Wilson coefficients are shown 
for the one, two, three, and four-dimensional SMEFT operator scenarios. 
It is found that 
the scenarios with modifications of right-handed $W$-$u$-$s$ couplings (by $C_{H ud}$) provide the best improvement relative to the SM and do not lead to problems in flavor physics or EWPO since constraints from $SU(2)_L$ invariance are not present. The scenarios with both left-handed and right-handed modifications display a slightly larger $\Delta \chi^2$, which can be understood by the fact left-handed operators change the EW fit by modifying $Z$-quark couplings.

\section{A heavy new physics interpretation: vector-like quark}

There have been several attempts to resolve the CAA by TeV-scale NP models \cite{Belfatto:2019swo,Coutinho:2019aiy,Cheung:2020vqm,Crivellin:2020lzu,Crivellin:2020ebi,Kirk:2020wdk,Crivellin:2021njn,Belfatto:2021jhf,Bryman:2021teu,Cirigliano:2022yyo,Crivellin:2022rhw,Belfatto:2023tbv,Endo:2020tkb,Capdevila:2020rrl,Li:2020wxi,Crivellin:2020oup,Crivellin:2020klg,Felkl:2021qdn,Branco:2021vhs,Marzocca:2021azj,Buras:2021btx,Crivellin:2021bkd,Cirigliano:2021yto,Blennow:2022yfm,Balkin:2022glu}. 
Among them, 
in this contribution, we consider vector-like quarks (VLQs) as they give rise to the preferred modifications at tree level. 
VLQs appear in many extensions of the SM such as grand unified theories~\cite{Hewett:1988xc,Langacker:1980js,delAguila:1982fs}, composite models or models with extra dimensions~\cite{Antoniadis:1990ew,Arkani-Hamed:1998cxo} and little Higgs models~\cite{Arkani-Hamed:2002ikv,Han:2003wu}. 
For the CAA tension,
 the effect of modified $Z$ couplings to quarks and loop effects in flavor observables have to be included in a global analysis.

There are seven possible VLQs mixed with quarks after EW symmetry breaking:
\begin{equation}
\begin{aligned}
U&: \QuantumNumbers{3}{1}{\frac{2}{3}} \,, &D &: \QuantumNumbers{3}{1}{-\frac{1}{3}} \,, &Q &: \QuantumNumbers{3}{2}{\frac{1}{6}} \,, \\
Q_5 &: \QuantumNumbers{3}{2}{-\frac{5}{6}} \,, &Q_7 &: \QuantumNumbers{3}{2}{\frac{7}{6}} \,, \\
T_1 &: \QuantumNumbers{3}{3}{-\frac{1}{3}} \,, &T_2 &: \QuantumNumbers{3}{3}{\frac{2}{3}} \,,
\end{aligned}
\end{equation}
with the representation under the SM gauge group $SU(3)\times SU(2)_L\times U(1)_Y$.
 The Lagrangian describing their interactions with the Higgs and SM quarks is 
\begin{align}
-\mathcal{L}_{\rm VLQ} =
&\xi_{fi}^{U}\bar{U}_{f}\tilde{H}^{\dagger}q_{i} 
+ \xi_{fi}^{D}\bar{D}_{f}H^{\dagger}q_{i} 
+\xi_{fi}^{u}\bar{Q}_{f}\tilde{H}u_{i} +\xi_{fi}^{d}\bar{Q}_{f}Hd_{i}
\\
+ &\xi_{fi}^{Q_{5}}\bar{Q}_{5,f}\tilde{H}d_{i}+\xi_{fi}^{Q_{7}}\bar{Q}_{7,f}Hu_{i} +\frac{1}{2}\xi_{fi}^{T_{1}}H^{\dagger}\tau\cdot\bar{T}_{1,f}q_{i}+\frac{1}{2}\xi_{fi}^{T_{2}}\tilde{H}^{\dagger}\tau\cdot\bar{T}_{2,f}q_{i} + \text{h.c.}\,,\nonumber
\end{align}
where $q$ is the left-handed quark doublet, $u,d$ are the right-handed quark singlets, and $i$ and $f$ are flavor indices for the SM quarks and new VLQs, respectively. Note that  $f$ does not necessarily need to run from 1 to 3 as the number of generations of VLQs is arbitrary.
 We disregard possible couplings between two VLQs representations and the SM Higgs as they are not relevant (at the dimension-six level) for the modification of gauge boson couplings to quarks.

With these conventions, the matching obtained by integrating out the VLQs at tree level onto the SMEFT is 
\begin{equation}
\label{eq:VLQ_SMEFT_tree_matching}
\begin{aligned}
\left[C_{Hu}\right]_{ij} &= -\frac{\xi^{u}_{fj} \xi^{u *}_{fi}}{2M_{Q_{f}}^2} + \frac{\xi^{Q_7}_{fj} \xi^{Q_7 *}_{fi}}{2M_{Q_{7f}}^2} \,, \\
\left[C_{Hd}\right]_{ij} &= \frac{\xi^{d}_{fj} \xi^{d *}_{fi}}{2M_{Q_{f}}^2} - \frac{\xi^{Q_5}_{fj} \xi^{Q_5 *}_{fi}}{2M_{Q_{5f}}^2} \,, \\
\left[C_{Hud}\right]_{ij} &= \frac{\xi^{d}_{fj} \xi^{u *}_{fi}}{M_{Q_{f}}^2} \,, \\
\CHqonebracket_{ij} &= \frac{\xi^{U}_{fj} \xi^{U*}_{fi}}{4M_{U_f}^2} - \frac{\xi^{D}_{fj} \xi^{D*}_{fi}}{4M_{D_f}^2} - \frac{3\xi^{T_1}_{fj} \xi^{T_1*}_{fi}}{16M_{T_{1f}}^2} + \frac{3\xi^{T_2}_{fj} \xi^{T_2*}_{fi}}{16M_{T_{2f}}^2} \,, \\
\CHqthreebracket_{ij} &= -\frac{\xi^{U}_{fj} \xi^{U*}_{fi}}{4M_{U_f}^2} - \frac{\xi^{D}_{fj} \xi^{D*}_{fi}}{4M_{D_f}^2} + \frac{\xi^{T_1}_{fj} \xi^{T_1*}_{fi}}{16M_{T_{1f}}^2} + \frac{\xi^{T_2}_{fj} \xi^{T_2*}_{fi}}{16M_{T_{2f}}^2} \,.
\end{aligned}
\end{equation}

\begin{figure}[t]
\centerline{\includegraphics[width=\textwidth]{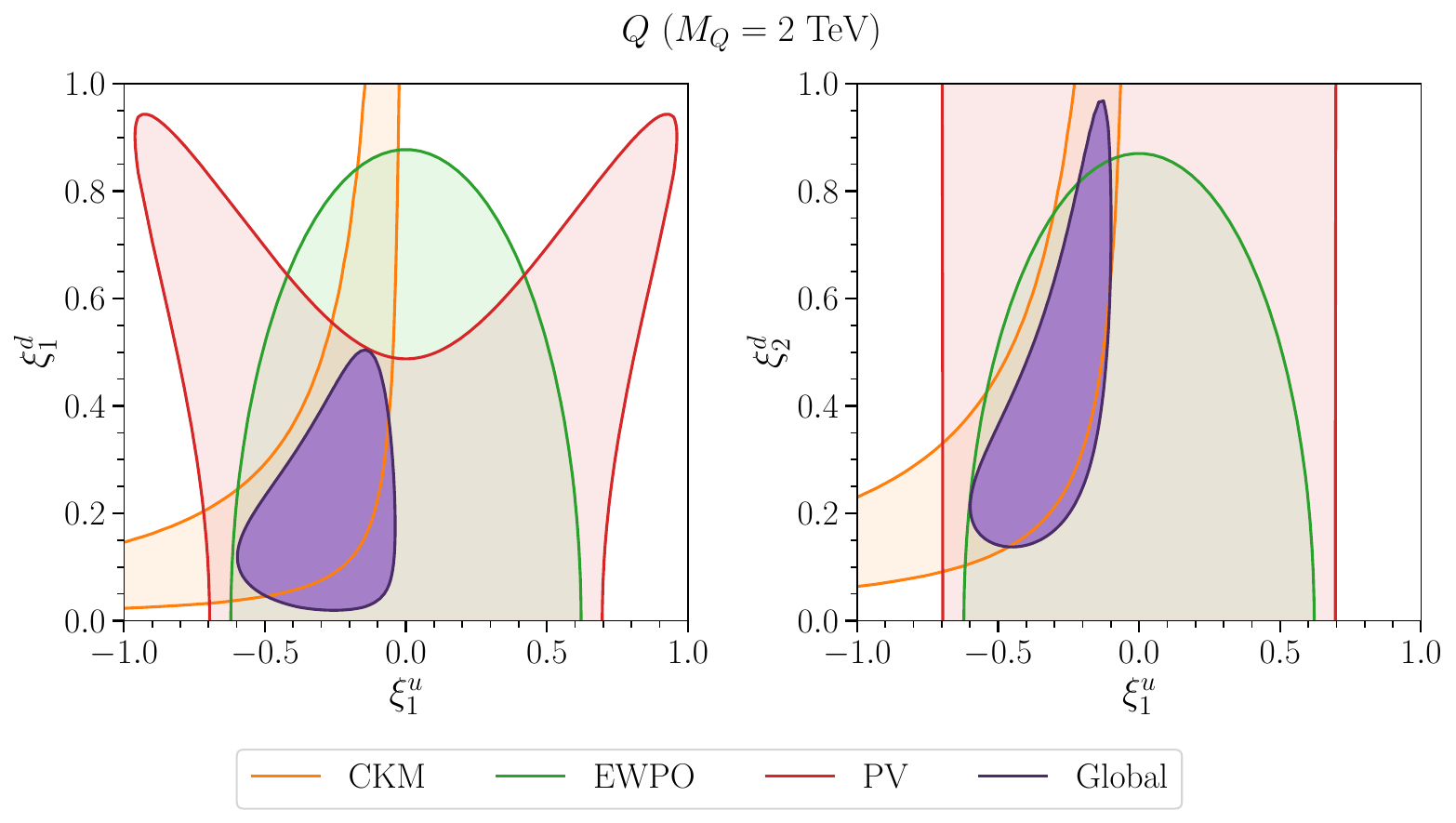}}
\caption{Global fits for the VLQ $Q$ with either couplings to $u$ and $d$ (left) or $u$ and $s$ scenario (right). \label{fig:Q}}
\end{figure}

The most important VLQ is the $SU(2)_L$ doublet $Q$, which is the only one that can contribute to the tight-handed $W$ coupling. 
For the $Q$ VLQ, since it can have both couplings to right-handed up and down quarks, we instead show two fits, for either purely $1^{\rm st}$ or purely $2^{\rm nd}$ generation down quark interactions in Fig.~\ref{fig:Q}. The study of the other VLQs is summarised in Ref.~\cite{Crivellin:2022rhw}.
 It is expected from our previous EFT results that there is a strong preference towards non-zero couplings to both right-handed $u$ and $d$ (left panel) or $u$ and $s$ (right panel).
However, unlike in the simple EFT scenario, the $Q$ field generates additional correlated effects in $Z$ couplings through $SU(2)_L$ invariance, and so parity violation (PO) and EWPO partially limit the parameter space. 
In the left panel, the best fit point is $\xi^u_1 = -0.29$, $\xi^d_1 = 0.21$ and has a pull w.r.t.\ the SM of  ${1.1}{\sigma}$. For the right panel, we find a best fit at $\xi^u_1 = -0.33$, $\xi^d_2 = 0.38$, and a pull of  ${2.1}{\sigma}$.
Furthermore, for a single generation of the doublet, NP in the right-handed $W$-$u$-$d$ and $W$-$u$-$s$ vertices at the same time lead to significant NP in right-handed $Z$-$d$-$s$ couplings, stringently constrained by $\epsilon_K$~\cite{Endo:2016tnu,Bobeth:2017xry}. 
Updating their result, we find that a single $Q$ doublet coupled to both $d$ and $s$ would have to obey $M_Q / \sqrt{\xi^d_1 \xi^d_2} >  {175}{\TeV}$ to be consistent with experiment, and therefore far too heavy to be relevant to the CAA.

Thus, a full explanation of the tensions in the Cabibbo angle determination require a modified $W$-$u$-$d$ and $W$-$u$-$s$ coupling and thus multiple generations of $Q$. Similarly, one can solve the CAA via a modified left-handed $W$-$u$-$d$ coupling and a right-handed $W$-$u$-$s$ coupling which again requires at least two VLQs. This means that a full solution of the CAA demands the presence of at least two VLQs.

\section{A light new physics interpretation: sterile neutrino}

Can the CAA be also explained by the light-scale new physics?
The one possibility is a MeV scale sterile neutrino scenario \cite{Kitahara:2023xab}.
In this section, we briefly summarize this scenario.

\begin{figure}[t]
\begin{center}
\includegraphics[width=0.8\textwidth]{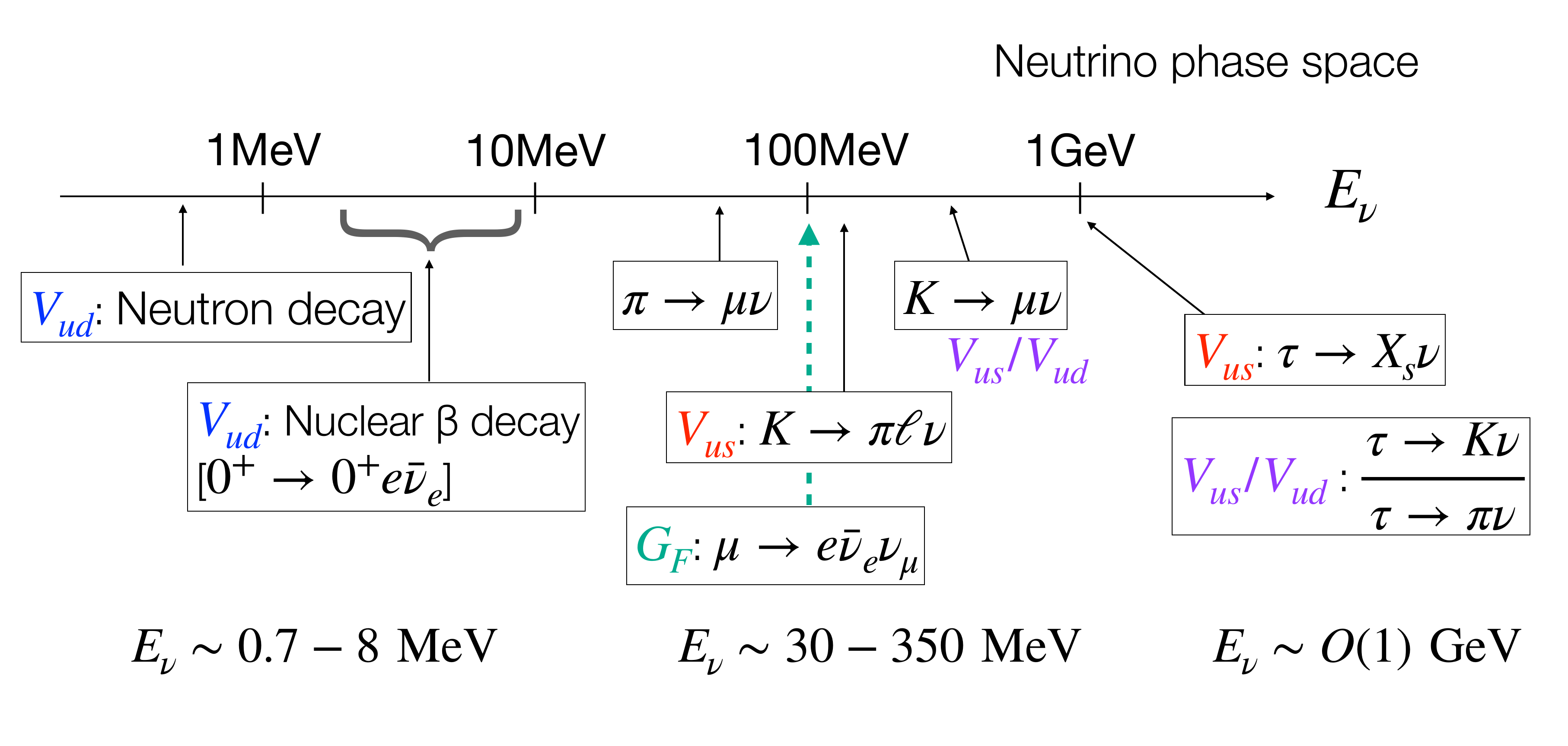}
\vspace{-10pt}
\caption{
Neutrino energy $E_\nu$ of each channel relevant to the first-row CKM unitarity test is summarized \cite{Kitahara:2023xab}. The sterile neutrino channel is open when $m_{\nu 4} < E_{\nu}$.
}\label{fig:Enu}
\end{center}
\end{figure}

Typically, a sterile neutrino's characteristics in most models can be effectively captured by its mass, $m_{\nu 4}$, and its mixing angles with the SM neutrinos, $U_{\ell 4}$.
Below the electroweak scale, the SM neutrinos from the weak doublets, $\nu_\ell$, split into two pieces in the mass eigenbasis, 
\begin{align}
\nu_\ell \simeq \cos U_{\ell 4} \nu_{\ell}^\prime + \sin U_{\ell 4}  \nu_{4} \quad \text{for~}\ell=e,\mu,\tau\,,
\label{eq:mixing}
\end{align}
where $\nu_{\ell}^\prime$ are the SM-like neutrinos, often called active neutrinos.
For the active neutrinos, the mixing ($\cos U_{\ell 4} \lesssim 1$) results in the coupling reduction of the weak interaction, and the deficit gives the sterile neutrino a feeble coupling ($\sin U_{\ell 4} \ll 1$) to the SM.   
These couplings alter observables primarily governed by the weak interaction, \ie, all the measurements relevant to the CAA are potentially affected, see Fig.~\ref{fig:Enu}. 
The sign of the modification depends on the mass scale of the sterile neutrino.

When the sterile neutrino mass exceeds $\mathcal O$(1)~GeV, 
it would not be kinematically permissible in the relevant measurements in Fig.~\ref{fig:Enu}. 
In this case, an important modification occurs in the Fermi-constant measurement via the muon decay. 
The observed value, $G_F^{\rm obs}$, deviates from the true value, $G_F$,  as described by
\begin{align}
G_F^{\rm obs} =G_F \cos U_{e 4}  \cos U_{\mu 4}\,. \label{eq:GF}
\end{align}
This relationship is pivotal in all the measurements employed in the CKM determinations.  
Using this relation, one can derive
\begin{align}
\Delta_{\rm CKM} = |V_{ud}^{\rm obs}|^2+|V_{us}^{\rm obs}|^2 -1 =\frac{|V_{ud}|^2}{\cos^2 U_{\mu 4}}+\frac{|V_{us}|^2}{\cos^2 U_{e 4}}-1 >0 \,. 
\end{align}
This is in contrast to the CAA in Eqs.~\eqref{eq:CAAglobal} and \eqref{eq:CAAglobalbeam}.

In the opposite limit, it is easily shown that when the sterile neutrinos are massless, all the measurements reproduce the SM predictions. 
 Hence, $\Delta_{\rm CKM}$ is expected to be zero,  which is again incompatible with the current data.

\begin{figure}[t]
\begin{center}
\includegraphics[width=0.48\textwidth]{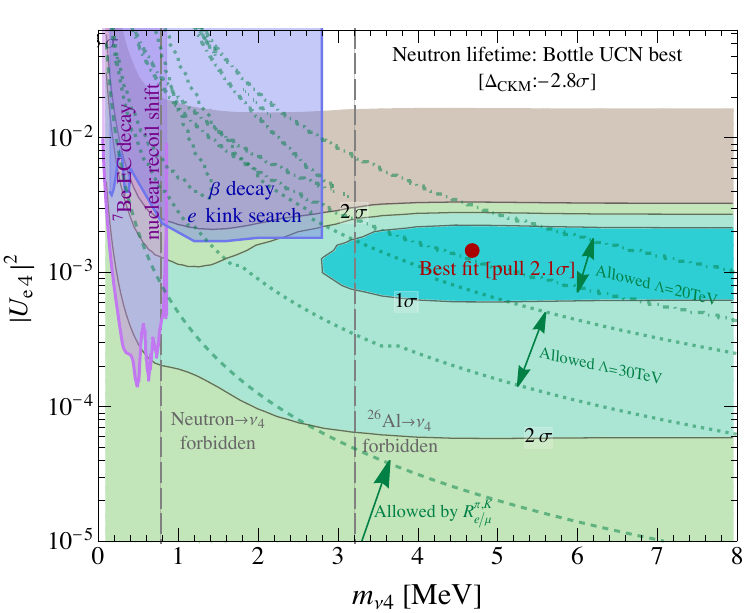} \quad 
\includegraphics[width=0.48\textwidth]{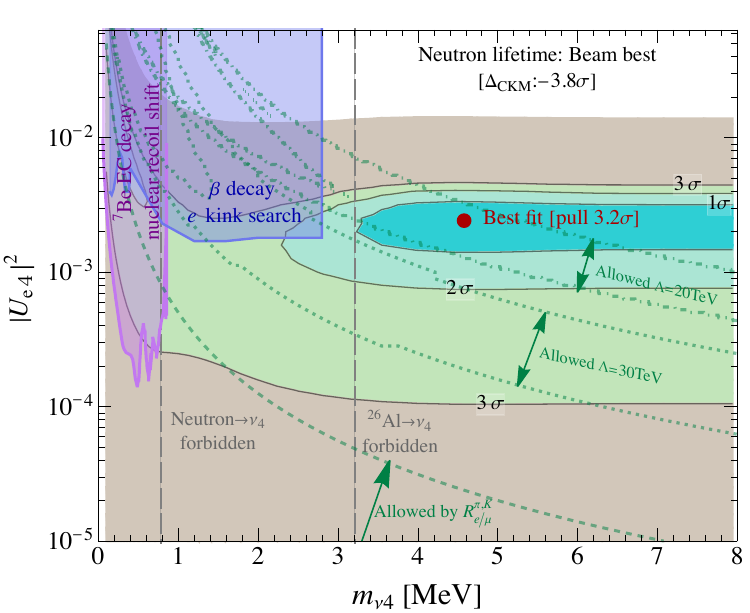}
\caption{Favored parameter regions are shown \cite{Kitahara:2023xab}.
The red points represent the best-fit point with the pulls from the SM hypothesis.
For the neutron decay data, 
the most precise neutron lifetime results from the bottle UCN and the beam measurements are used in the left and right panels, respectively. 
The blue and purple shaded regions are excluded by the nuclear $\beta$-decay kink searches  \cite{Schreckenbach:1983cg,Deutsch:1990ut,Derbin:2018dbu,Bolton:2019pcu} and the EC-decay search  \cite{Friedrich:2020nze}, respectively.
The regions above green dashed lines can be constrained by $\pi^+ \to e^+ \nu$ measurements \cite{PiENu:2015seu,Bryman:2019ssi,Bryman:2019bjg}, and the regions between dotted or dash-dotted green lines are allowed with the dimension-six operator. See more details in the text.
\label{fig:neutrinofit}}
\end{center}
\end{figure}

Interestingly, when the sterile neutrino mass is in the MeV scale,
 the coupling reduction of the weak interaction remains in the neutron and nuclear decays, while other observables, especially the Fermi constant, stay almost the same as in the SM.
 Consequently, only $V_{ud}$ is modified  \cite{Kitahara:2023xab},
\begin{align}
    |V_{ud}^{\rm obs}|^2
    = \left[
    1- \epsilon(m_{\nu 4}, \delta M)\sin^2 U_{e4}\right] |V_{ud}|^2\,,
\end{align}
with
\begin{align}
\epsilon\left(m_{\nu 4}, \delta M\right)\equiv 1-\frac{I(m_{\nu 4}, \delta M)}{I(0, \delta M)}\,,
\end{align}
and
\begin{align}
I\left(m_\nu, \delta M\right) \equiv \int_{m_e}^{\delta M-m_\nu} dE_e \sqrt{E_e^{2}-m_e^{2}}\sqrt{E_\nu^{2} -m_\nu^{2}} E_e E_\nu\,, 
\end{align}
where   $\delta M =  Q_{\rm EC}-m_e$ and the  electron-capture (EC) $Q$-value is  $Q_{\rm EC}$.
The other quantities, $G_F^{\rm obs}, V_{us}^{\rm obs}$, and $(V_{us}/V_{ud})^{\rm obs}$, are the same as the SM ones because the corresponding $E_\nu$ is much larger than MeV. This realizes the experimentally favored value,
$
\Delta_{\rm CKM} <0   
$.
From the size of the anomaly $\Delta_{\rm CKM} \approx 10^{-3}$,
one can infer that the favored mixing-angle-squared is $U^2_{e 4}\approx 10^{-3}$, and $U^2_{\mu 4}$ is not necessary.

The global analyses are shown in Fig.~\ref{fig:neutrinofit}. 
We show two panels depending on the neutron lifetime measurements. In the left panel,  
we use the bottle measurement to extract $V_{ud}$, which is consistent with the one from the super-allowed nuclear decays. The original tension of the CAA is at 2.8$\sigma$, and the pull in the presence of sterile neutrino is 2.1$\sigma$ at the best-fit point.
On the other hand, since the beam measurement prefers smaller $V_{ud}$ than the super-allowed nuclear decays, 
a significance of the CAA is enhanced to be 3.8$\sigma$, and the sterile neutrino can relax it by  3.2$\sigma$ at the best-fit point. 

This mass region of the sterile neutrino receives several constraints from the laboratory to the cosmology. All constraints are investigated in Ref.~\cite{Kitahara:2023xab}, and it is shown that 
there are viable scenarios within an extension of the inverse seesaw model.

\section{Conclusions}

Improvements in lattice results for the kaon form factors and also the radiative corrections
have revealed a mild tension in the first-row CKM unitarity test. 
The right-handed $W$ couplings are preferred in light of the tension, and the prime candidate for the UV completion is the vector-like quark extension. 
Explanation by a MeV sterile neutrino is also possible, although the viable model is challenging.
The STCF is needed for the first-column and second-row unitarity tests.

\section*{Acknowledgments}
I would like to thank Andreas Crivellin, Matthew Kirk, Federico Mescia, and Kohsaku Tobioka for fruitful collaborations on the presented works.
The presented works were supported by the Grant-in-Aid for Early-Career Scientists from the Ministry of Education, Culture, Sports, Science, and Technology (MEXT), Japan, No.\,19K14706 and by the Japan Society for the Promotion of Science (JSPS)  Core-to-Core Program, No.\,JPJSCCA20200002.
 I also want to warmly thank the organizers of the 2024 International Workshop on Future Tau Charm Facilities for inviting me and giving me the opportunity to present these results at this great conference.

\bibliographystyle{ws-ijmpa}
\bibliography{ref}
\end{document}